\documentclass[twocolumn]{IEEEtran} 

\usepackage[dvips]{graphicx}

\def\BibTeX{{\rm B\kern-.05em{\sc i\kern-.025em b}\kern-.08em
    T\kern-.1667em\lower.7ex\hbox{E}\kern-.125emX}}

\begin{document}

\title{Utilizing Reconfigurable Hardware Processors via Grid Services} 

\author{Darran Nathan, Ralf Clemens\thanks{The authors are with the
DSP Technology Centre, School of Engineering, NgeeAnn Polytechnic, Singapore. (e-mail: darran@projectproteus.org [Darran Nathan]).}}

\markboth{Project Proteus}
{Murray and Balemi: Using the Document Class IEEEtran.cls} 

\maketitle

\begin{abstract}
Computational grids typically consist of nodes utilizing ordinary processors such as the Intel Pentium. Field Programmable Gate Arrays (FPGAs) are able to perform certain compute-intensive tasks very well due to their inherent parallel architecture, often resulting in orders of magnitude speedups. This paper explores how FPGAs can be transparently exposed for remote use via grid services, by integrating the Proteus Software Platform with the Globus Toolkit 3.0.
\end{abstract}

\begin{keywords}
reconfigurable computing, grid computing
\end{keywords}

\section{Introduction}
\label{sectIntroduction}
Grid Computing describes an advanced distributed computing infrastructure, where resources are dynamically allocated according to the processing task requirements. These resources are usually compute nodes that may be located within an organization, or may be distributed across different organizations, in different geographical locations. To tie all these resources together, various grid computing 'middleware' have been developed. Among the more popular is the Globus Toolkit \cite{bibGlobus}, which also offers an API that programmers can use to easily expose program functionality as a 'Grid Service'.

Compute-intensive tasks have traditionally been run on grids / clusters of nodes having ordinary processors such as the Intel Pentium. More recently, there has been an increasing interest in the use of reconfigurable hardware processors called Field Programmable Gate Arrays (FPGAs) to undertake these processing tasks. The inherent parallelism of processing done on FPGAs often allows for orders of magnitude speedups \cite{bibSW} compared to a normal processor.

This paper explores how FPGAs can be transparently exposed for remote use via a grid service, by utilizing the Globus Toolkit and the Proteus Software Platform (PSP) \cite{bibPSP}. Section \ref{sectPSP} describes the architecture of the PSP and how it abstracts the underlying FPGA, Section \ref{sectGlobus} discusses the architecture of the Globus middleware, Section \ref{sectIntegration} presents how the PSP was linked with Globus to allow for use of an FPGA via a Grid Service, and finally Section \ref{sectConclusion} concludes the paper.

\section{The Proteus Software Platform}
\label{sectPSP}
\begin{figure}[htb]
\begin{center}
\includegraphics[width=0.2\textwidth]{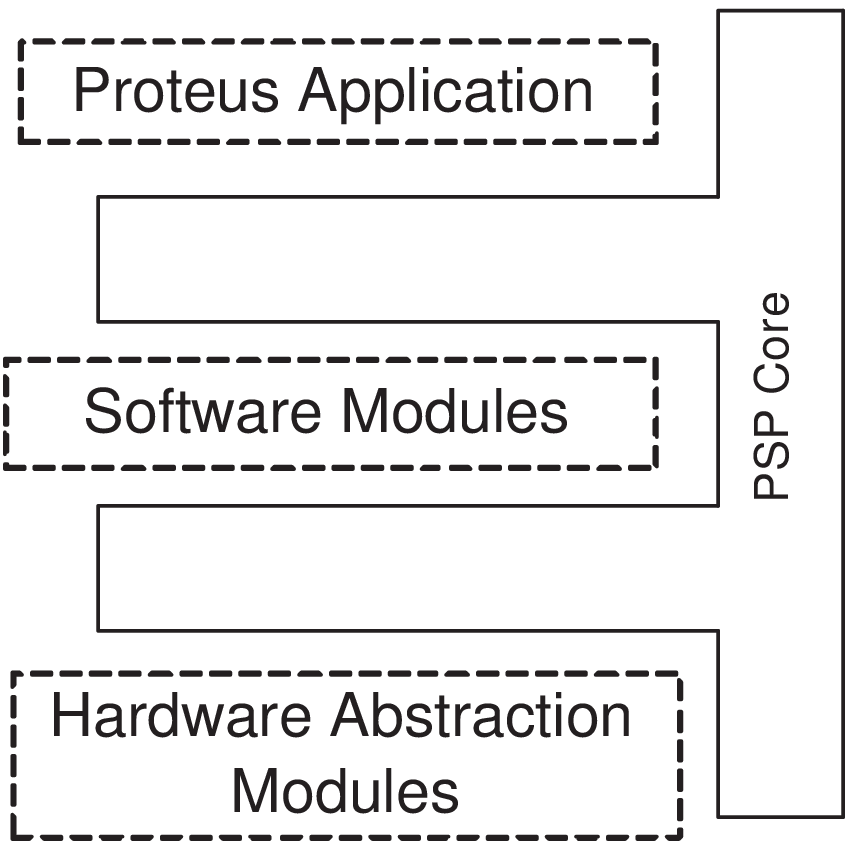}
\caption{Components of the PSP}
\label{figPSPComponents}
\end{center}
\end{figure}

The Proteus Software Platform (PSP) can be divided into four main component blocks: the PSP core, which holds the common set of interfaces and functionality, and three other components - the Software Modules, the Hardware Abstraction Modules (HAMs), and the Proteus Application. This segmentation is illustrated in Figure \ref{figPSPComponents}.

\begin{figure}[htb]
\begin{center}
\includegraphics[width=0.5\textwidth]{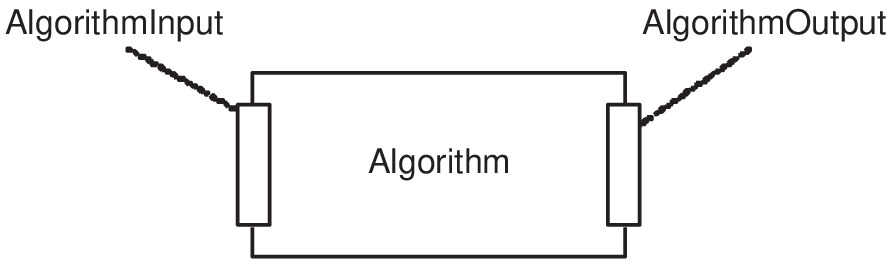}
\caption{Algorithm block representation}
\label{figAlgo}
\end{center}
\end{figure}

A Software Module consists of a set of Algorithm blocks which are linked up to form the desired processing chain. Each block defines a unit of operation that receives data at an input, processes it, and sends the results out through an output. This is commonly represented as shown in Figure \ref{figAlgo}.

\begin{figure}[htb]
\begin{center}
\includegraphics[width=0.5\textwidth]{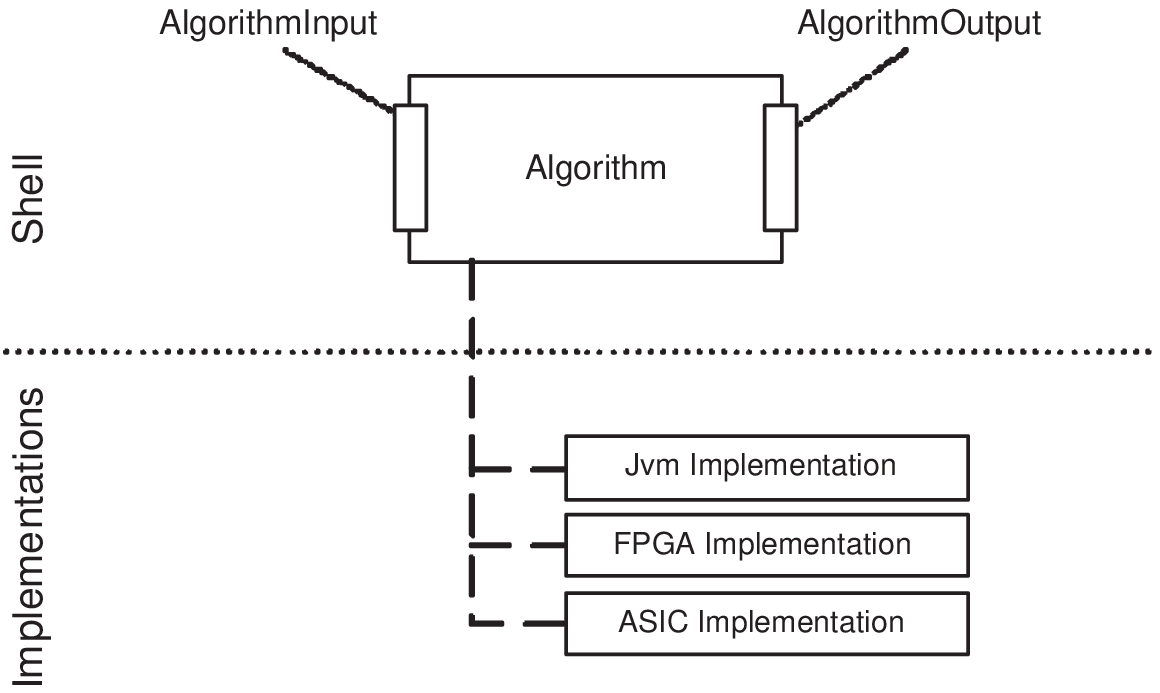}
\caption{The Algorithm 'shell / implementation' structure}
\label{figShellImp}
\end{center}
\end{figure}

Since the PSP is intended to be run in environments where the available processor types are variable and determined only during run-time, each Algorithm block may be associated with a number of implementations for different processor types. The resulting design takes on a 'shell / implementation' architecture as shown in Figure \ref{figShellImp}.

\begin{figure}[htb]
\begin{center}
\includegraphics[width=0.5\textwidth]{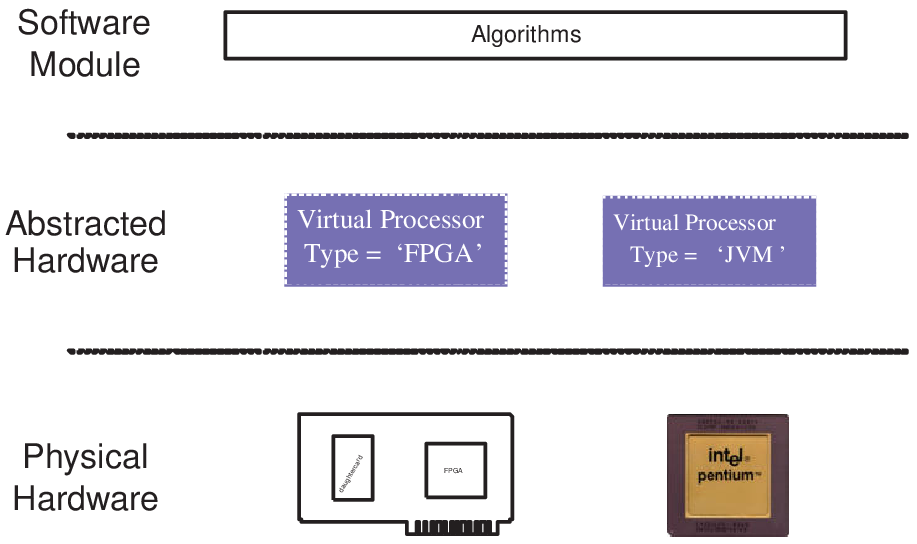}
\caption{Abstraction of physical hardware via 'virtual processors'}
\label{figHAM}
\end{center}
\end{figure}

A Hardware Abstraction Module (HAM) defines a layer of abstraction to the underlying hardware, allowing the PSP to access and utilize various kinds of processor types and other PC resources via a common interface. This is done by modelling the desired properties of one or more physical hardware resources in one or more 'virtual processor' entities, as shown in Figure \ref{figHAM}. The PSP dynamically matches the available algorithm implementations with the available virtual processors, deploying them accordingly to build up the processing chain described by the Algorithm shells.

The final component, the Proteus Application, defines the interface to the end-user.

\section{The Globus Toolkit}
\label{sectGlobus}
The Globus Toolkit 3.0 exposes an API which programmers can utilize to develop grid services as well as clients that access these services. In creating a grid service, the supported calls are specified in its associated WSDL file. The exchange of this file and the underlying connection establishment protocol between the client and the grid service is encapsulated by the toolkit, and is therefore transparent to the programmer.

The PSP and the Globus Toolkit are both built on the Java \cite{bibJava} platform, hence simplifying the integration work needed for the PSP to sink / source data via a grid service. The next section describes in greater detail how this has been done.

\section{Exposing use of the FPGA via a Grid Service}
\label{sectIntegration}
\begin{figure}[htb]
\begin{center}
\includegraphics[width=0.4\textwidth]{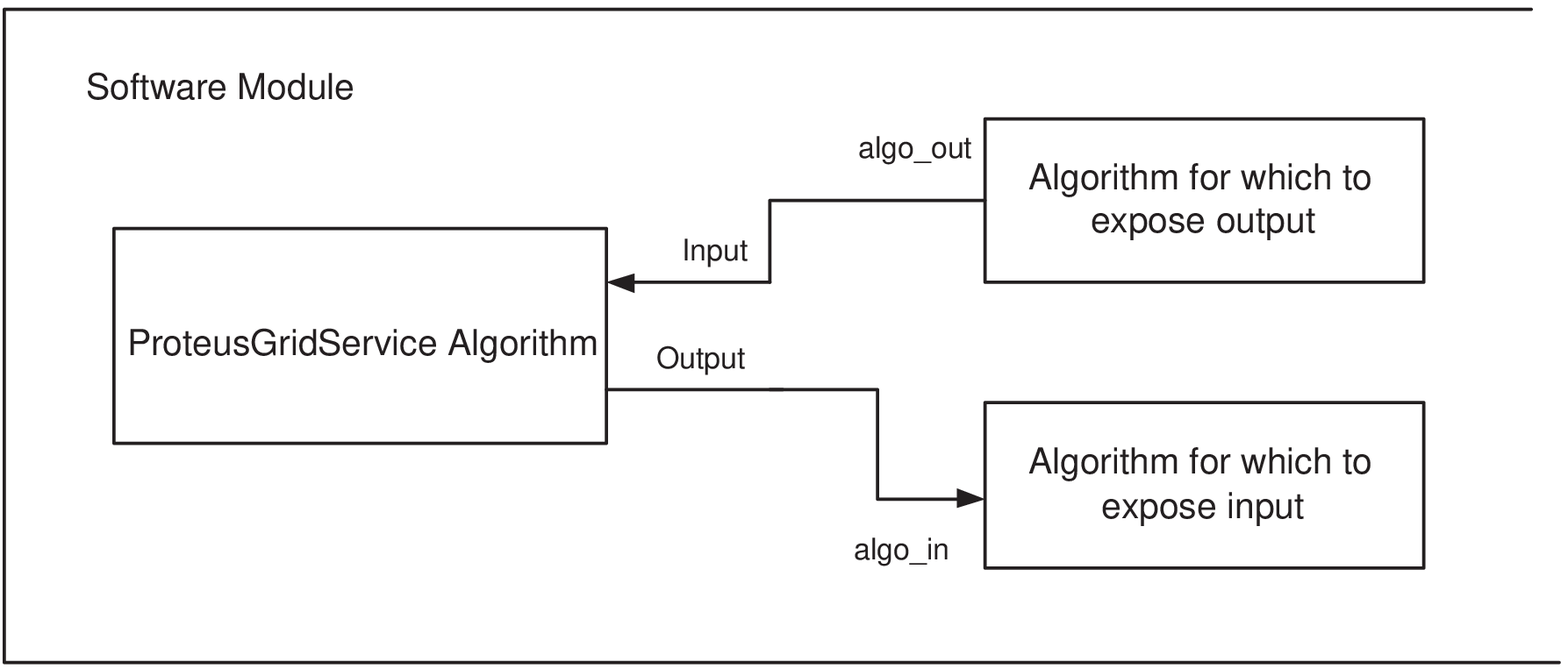}
\caption{Software Module with ProteusGridService Algorithm}
\label{figGridSoftwareModule}
\end{center}
\end{figure}

A Globus HAM has been developed to represent a 'grid' type virtual processor, which encapsulates calls to exchange data via the Globus Toolkit. A 'ProteusGridService' Algorithm block compatible with this virtual processor type has been developed, and can be linked up to the rest of the Algorithm chain of the Software Module, as shown in Figure \ref{figGridSoftwareModule}.

\begin{figure}[htb]
\begin{center}
\includegraphics[width=0.5\textwidth]{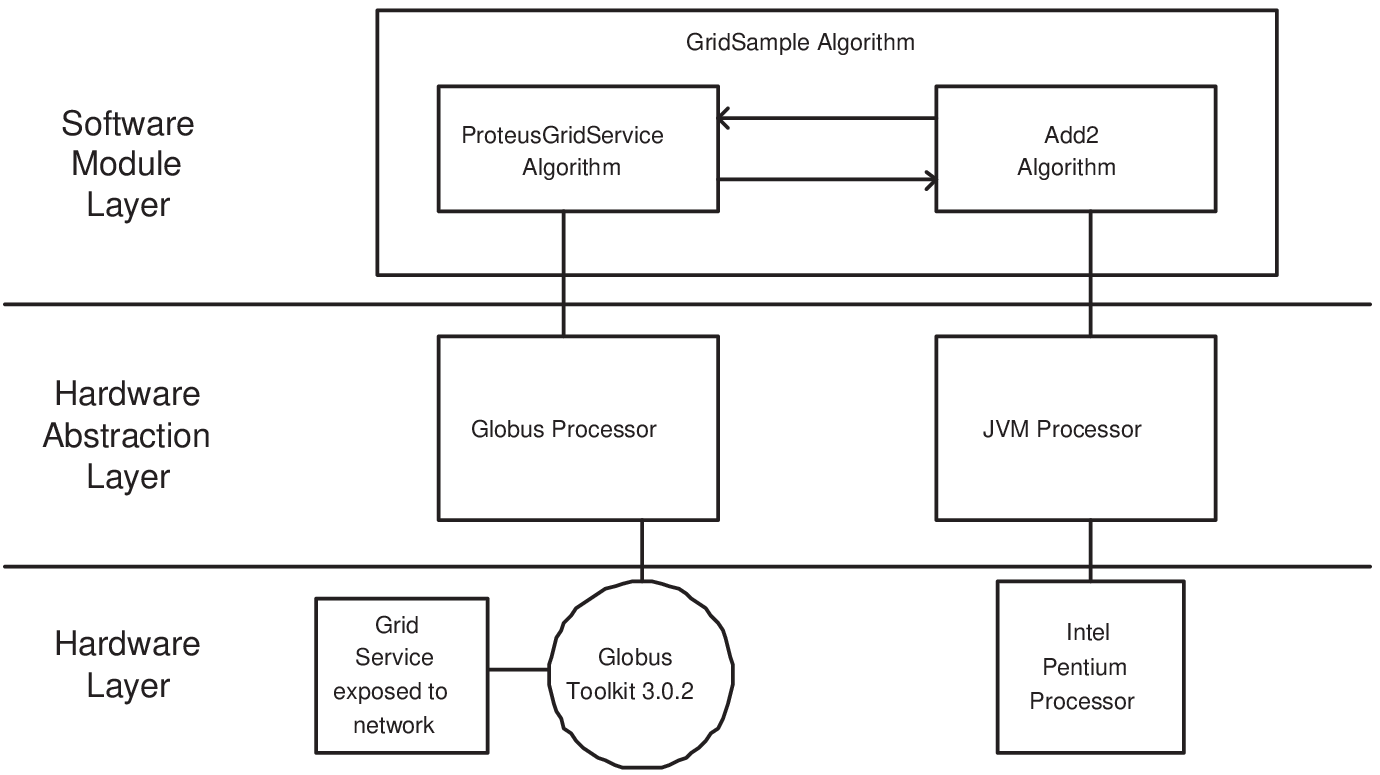}
\caption{Layers of data exchange}
\label{figLayers}
\end{center}
\end{figure}

Data to be sent / retrieved via the grid service is passed through the 'ProteusGridService' Algorithm, the Globus HAM's virtual processor, and from there through the Globus Toolkit. These layers of data exchange is illustrated in Figure \ref{figLayers}.

\begin{figure}[htb]
\begin{center}
\includegraphics[width=0.4\textwidth]{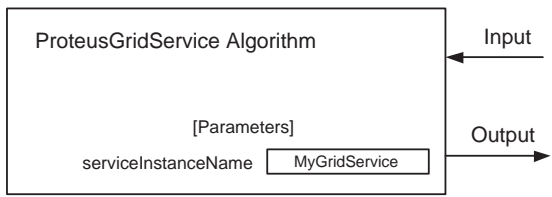}
\caption{ProteusGridService Algorithm}
\label{figProteusGridServiceAlgo}
\end{center}
\end{figure}

The 'ProteusGridService' Algorithm block has a 'serviceInstanceName' parameter that is initialized by the Software Module. This name will be used by the Globus virtual processor as that of the grid service name associated with this Software Module. Figure \ref{figProteusGridServiceAlgo} shows a ProteusGridService Algorithm instantiated with the 'serviceInstanceName' parameter initialized to "MyGridService".

When the PSP deploys this Software Module, the ProteusGridService Algorithm will be attached to the Globus virtual processor. Starting this virtual processor will invoke the embedded Globus service container to expose a grid service of name "MyGridService". The full grid service URL will correspondingly be:

\small
\begin{verbatim}
http://127.0.0.1:8080/ogsa/services/
proteusgrid/ProteusGridService/MyGridService
\end{verbatim}
\normalsize

\begin{figure}[htb]
\begin{center}
\includegraphics[width=0.5\textwidth]{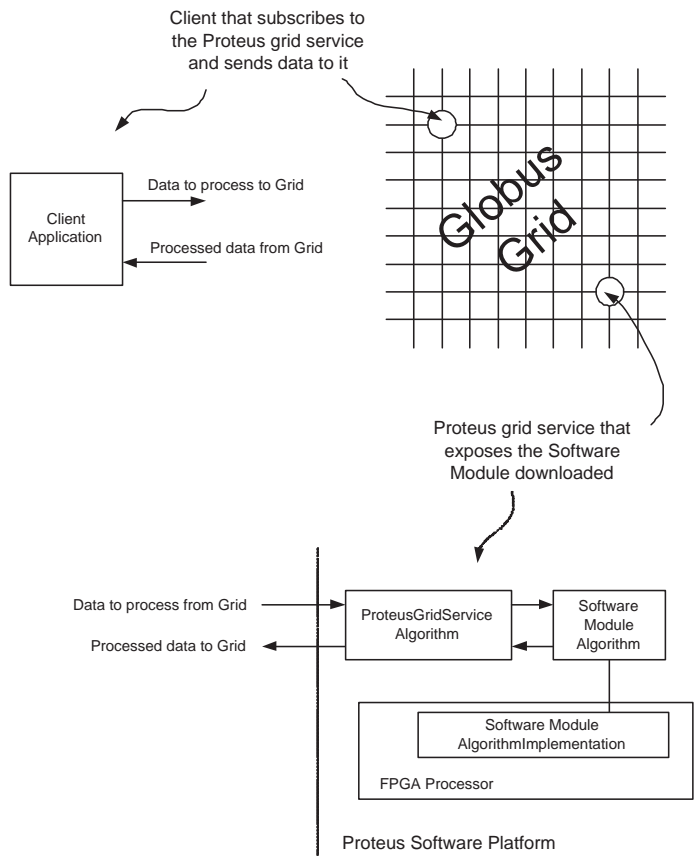}
\caption{Remote client access of grid service}
\label{figRemoteClient}
\end{center}
\end{figure}

Once started, any remote client can subscribe to the grid service and exchange data with it, as shown in Figure \ref{figRemoteClient}.

\section{Conclusion}
\label{sectConclusion}
This paper has presented how FPGAs can be exposed for remote use via grid services, by integrating the Proteus Software Platform with the Globus Toolkit 3.0.

\section*{Acknowledgments}
We gratefully acknowledge the funding support provided by the Ngee Ann Kongsi (Singapore) and Ngee Ann Polytechnic's Innovation \& Enterprise Office. Special thanks to the rest of the Proteus Team - Philip Wong, Kelvin Lim, Andreas Weisensee, and Kelly Choo.

\nocite{*}
\bibliographystyle{IEEE}

%

%

\end{document}